# Investigation of laser-induced-metal phase of MoTe$_2$ and its contact property via scanning gate microscopy

Kohei Sakanashi,[1,*] Hidemitsu Ouchi,[1] Kota Kamiya,[1] Peter Krüger,[1,2] Katsuhiko Miyamoto,[1,2] Takashige Omatsu,[1,2] Keiji Ueno,[3] Kenji Watanabe,[4] Takashi Taniguchi,[4] Jonathan P. Bird,[1,5] and Nobuyuki Aoki[1,2,*]

[1]Department of Materials Science, Chiba University, Chiba, 263-8522, Japan

[2]Molecular Chirality Research Center, Chiba University, Chiba, 263-8522, Japan

[3]Department of Chemistry, Saitama University, Saitama, 338-8570, Japan

[4]Advanced Materials Laboratory, National Institute for Materials Science, Tsukuba, 305-0044, Japan

[5]Department of Electrical Engineering, the State University of New York, University at Buffalo, Buffalo, New York, 14260, USA

E-mail: k.sakanashi@chiba-u.jp, n-aoki@faculty.chiba-u.jp

**Abstract**

Although semiconductor to metal phase transformation of MoTe$_2$ by high-density laser irradiation of more than 0.3 MW/cm$^2$ has been reported, we reveal that the laser-induced-metal (LIM) phase is not the 1T′ structure derived by a polymorphic-structural phase transition but consists instead of semi-metallic Te induced by photo-thermal decomposition of MoTe$_2$. The technique is used to fabricate a field effect transistor with a Pd/2H-MoTe$_2$/LIM structure having an asymmetric metallic contact, and its contact properties are studied via scanning gate microscopy. We confirm that a Schottky barrier (a diffusion potential) is always formed at the Pd/2H-MoTe$_2$ boundary and obstacles a carrier transport while an Ohmic contact is realized at the 2H-MoTe$_2$/LIM phase junction for both n- and p-type carriers.

**1. Introduction**

Transition metal dichalcogenides (TMDs) are receiving a large amount of attention because of their rich physical properties, such as their valley degree of freedom[1–3] and polymorphism[4–6], and the potential that these offer for use in future flexible field effect transistor (FET) devices. However, achieving Ohmic contact to semiconducting TMDs materials is very difficult, due to Fermi level pinning[7] and the van der Waals (vdW) gap between the semiconducting channel and metal electrodes.[8] Experimentally the Schottky barrier height has been studied precisely for conventional metals and the pinning factor has been derived as ~0.1 for MoS$_2$.[9] Formation of edge contact is a promising method to realize low contact resistance for graphene device[10], however the technique has not been well established for TMDs yet. A graphene contact method[11,12] or monolayer hexagonal-boron nitride (*h*-BN) tunneling layer contact method[13] have been developed for realization of Ohmic contact to TMDs even for low temperatures, but they are quite complex and highly technical methods. Therefore, such techniques are not suitable for commercial device application.

MoTe$_2$ is a TMD which shows interesting polymorphism[14–16]; the 2H and the 1T (or 1T′) phase are semiconducting and semi-metallic, respectively. Among the TMDs, MoTe$_2$ has a relatively small band gap. The band gaps of monolayer and multilayer MoTe$_2$ are 1.1 eV and 0.88 eV[17], respectively, comparable to that of bulk silicon. Because of the narrow gap, 2H-MoTe$_2$ sometimes shows ambipolar-type FET properties.[17,18] Recently, by irradiation with a strong continuous-wave (CW) green-laser, it has been reported that a phase transition occurs from the 2H semiconducting to the 1T′ semi-metallic phase.[19–21] As a result, the carrier injection properties from the laser-induced-metal (LIM) phase[19] or a real-1T′ crystal[22] to the 2H semiconducting one are improved as compared to the usual metal/2H contact because of the relatively small potential barrier height between the 2H and 1T′ interface. Since a finite Schottky barrier height still remains at the 2H/1T′ interface, the use of 1T′ contact cannot solve this critical issue.[22]

Nonetheless, it is questionable whether the LIM phase is really composed of 1T′-MoTe$_2$, because the Raman spectrum of the LIM phase is very different from that of a thermally grown 1T′-MoTe$_2$ crystal, as we have found. Moreover, there is still room for discussion whether the LIM contact provides effective contact to both n- and p-type carriers, or not. In this paper, we fabricated a FET sample comprised of the Pd/2H-MoTe$_2$/LIM structure and studied the contact properties using scanning gate microscopy (SGM); this can visualize barrier formation[23,24], quantum interference effects[25], carrier trajectories[26,27], and so on, by using a conductive AFM tip as a movable point gate electrode for local electrostatic perturbation.

## 2. Device Fabrication and Experimental Procedure

All samples used in this paper were fabricated by using chemical-vapor-transport (CVT)-grown multi-layer 2H-MoTe$_2$ flakes[28] directly exfoliated onto thermally grown 300 nm SiO$_2$/p$^{++}$-Si substrates. The LIM phase was fabricated in 2H-MoTe$_2$ by irradiation of a CW green laser ($\lambda$ = 532 nm), focused by 100× objective lens at the sample surface to achieve an optical density of more than 1 MW/cm$^2$. The phase transition from 2H to LIM occurs within one second (see supporting information #S1 for details). The devices A and B shown in Figs. 1(a) and 1(b), used in the electrical measurements, were fabricated by standard electron-beam lithography and electron-beam metal deposition; the electrical contacts were provided by Cr (5 nm)/Au (80 nm). In device B, a line of LIM region was drawn across the channel. The device C used in the SGM observation was fabricated by drawing some lines of LIM before the deposition of metallic electrodes of Pd (5 nm)/Au (20 nm). The electric contacts at the left and the right side are achieved by Pd and LIM lines, respectively. Raman spectra and mapping were measured in a JASCO RPM-510 and NRS-7100 Raman microspectrometer, respectively, with an excitation wavelength of 532 nm and a 100× objective lens. In order to avoid undesirable thermal effects, the laser power was kept below 1 mW in these experiments. Our ambient SGM measurement was built as custom units based on commercial scanning probe microscope (PicoPlus$^{TM}$, Molecular Imaging,[23] see supporting information #S2 for details). During every SGM observation, a tip-voltage ($V_T$) of 5 V was applied to a commercial Pt/Ir-coated AFM cantilever and the tip was lifted 100 nm above the device using the interleave mode. Note that our SGM set-up uses a laser-based AFM system but the channel length is smaller than the cantilever, and so the semiconductor channel region is masked by the cantilever. Therefore, photo induced effect for 2H-MoTe$_2$ is avoided. All the electrical measurements, Raman spectroscopy and SGM observation were performed at ambient condition.

## 3. Results and discussion

### 3.1. Strong laser-induced phase transition

The phase transition from 2H structure to LIM phase occurs relatively easily, simply by using an objective lens to irradiate the CW green laser (power density of more than 0.3 MW/cm$^2$) in air, and so it can be used for direct wiring of metallic region in a single 2H-MoTe$_2$ crystal as shown in Fig. 1(b). The laser-irradiated region becomes thinner and there remains an approximately 10-nm thick LIM phase. Surprisingly, and in contrast to previous research[15,17], the Raman spectrum of the LIM phase is very different from that of an exfoliated real 1T′ crystal, as shown in Fig. 2, but it is quite similar to the Raman spectra of pure Te. This huge difference in the Raman spectrum, between the LIM phase and a real 1T′ crystal indicates that the strong laser irradiation has decomposed MoTe$_2$ into Mo and Te rather than causing a structural phase transition from 2H to 1T′.[19] The same LIM phase transition was confirmed even when such a strong laser was irradiated onto a 1T′-MoTe$_2$ crystal as shown in Fig. 2. Small traces of Raman peaks related to MoO$_x$ are observed, but the identification of the critical chemical composition is difficult[29]. However, it is reasonable to consider that MoO$_x$ does not contribute to the metallic behavior of the LIM phase since the MoO$_x$ is usually an insulator.[30] The existence of Mo atoms is confirmed also by energy dispersive X-ray spectroscopy (EDS, see supporting information #S3 for the detail). Interestingly, the ratio of Mo and Te is almost 1:1 in the LIM region. Since the vapor pressure of Te is higher than that of Mo, Te could easily be vaporized during the photo-thermal decomposition as discussed later. A possible reason why no trace of Mo is observed in the Raman spectra is that Mo can be oxidized suddenly under ambient condition. We note that thermal decomposition of $MoTe_2$ into $MoO_x$ and Te has been reported as being caused by rapid annealing and cooling of $2H-MoTe_2$ in ambient.[31] In our laser irradiation system, a 50-mW green laser is focused by a 100× objective lens. From this, and the temperature dependence of Raman peak shift of hexagonal boron-nitride (h-BN) on a $MoTe_2$ flake (See supporting information #S4)[32], it can be estimated that during the LIM transition, the temperature increases to about 1250 K, which is high enough to decompose the $MoTe_2$ crystal. Interestingly, the Raman spectrum of the LIM region indicates the existence of non-oxidized Te and the sample shows good metallic behavior. In fact, in device B, which has a line of LIM region in the channel, we observed the gate voltage ($V_{BG}$) independent behavior while the conventional 2H channel device (device A) showed ambipolar-semiconducting behavior as shown in Fig. 1(c). The conductivity of the LIM channel is more than two orders of magnitude larger than that of the ON state of the 2H phase FET. Although single crystal Te is generally considered a very-narrow-gap semiconductor[33], our density-functional theory (DFT) calculations indicate the Te can have a semi-metallic band structure when spin-orbit coupling is accounted for (See supporting information #S5). Moreover, a metallic temperature dependence of a LIM channel has been confirmed in the other sample which has a similar structure as device B (See supporting information #S6). The resistance went down when decreasing the temperature to 300 mK. No gate-voltage dependence of the resistance was observed even at low temperature.

### 3.2. Ohmic carrier injection from LIM electrode

In order to confirm whether the LIM contact is effective as an Ohmic contact to both n- and p-type $MoTe_2$, we have prepared the device C as shown in Fig. 3(a). Transition from the $2H-MoTe_2$ to the LIM phase was confirmed by Raman spectrum mapping around the channel region superimposed to the optical image as shown in Fig. 3(b). The region colored in green corresponds to a Raman peak at 120 cm$^{-1}$, which is the main peak of both the LIM phase and Te. This region is situated along the laser-drawn lines on the MoTe$_2$ crystal, except for those regions underneath the top-contacted Pd/Au electrode. The distribution of the green color region is slightly wider than the width of the LIM region. As can be confirmed from the line-profile of an atomic force microscope (AFM) image shown in Fig. 3(c), the thickness of the laser irradiated region is approximately 10 nm which is thinner than that of the pristine MoTe$_2$ crystal (~30 nm). Therefore, the laser-decomposed materials, including Te, were flushed from the laser-irradiated region and redeposited on the crystal along the laser-irradiated lines. Fig. 3(d) shows the $V_{BG}$ dependence of $I_{SD}$ for different bias conditions where the Pd contact (the left side of the channel in Fig. 3(a)) was grounded and the LIM one is positively or negatively biased ($V_{SD}$). The red curve shows the $I_{SD}$ behavior when a voltage of −0.1 V was applied at the LIM electrode. Both negative and positive carriers can be injected from the LIM and the Pd contacts, respectively, into the MoTe$_2$ channel, showing an ambipolar behavior by sweeping the $V_{BG}$ from negative to positive. On the other hand, when +0.1 V is applied at the LIM electrode, positive charges can be injected from the LIM contact, but it is hard to inject negative charges from the Pd contact, resulting in p-type behavior shown as the blue curve. Such characteristics can also be confirmed in the $I_{SD}$-$V_{SD}$ curves as shown in Figs. 3(e) and 3(f). For positive carriers at $V_{BG}$ = −40 V, the holes can be injected from both sides of the channel as shown in Fig. 3(e). However, Fig. 3(f) shows a diode-like rectification for negative carriers; electron injection is strongly suppressed at the Pd contact due to the existence of a 20 ~ 80 meV-high Schottky barrier at the interface between Pd and a valence band of 2H-MoTe$_2$ extracted from several temperature-dependence experiments.[34–36] Considering a Fermi-level pinning effect[30], the barrier height from the Fermi level of Pd to the conduction band of 2H-MoTe$_2$ is much larger than that to the valence band, because of the larger work function of Pd (~5.4 eV).[37]

Such a diode-like behavior has been reported in a MoS$_2$ device due to asymmetric Schottky contact.[38] However, in device C, the contact at MoTe$_2$/LIM interface is considered as Ohmic since no trace of Schottky barrier formation was observed in the SGM measurement as discussed later. As for the reason why an Ohmic contact forms at the LIM, several mechanisms can be considered. The LIM phase may have a vanishing van der Waals gap with the 2H-MoTe$_2$ crystal due to the creation of in-plane (hetero) junction of LIM and 2H phase. Also, due to the semi-metallic band structure of Te, it can be expected that Fermi level pinning effect between the MoTe$_2$ and the LIM phase does not occur, as in the case of a graphene contact used in a recent study.[11] Moreover, the strong laser irradiation induces a high-density carrier doping in the MoTe$_2$ crystal around the LIM region because chalcogen atoms, especially Te, are easily displaced from the crystal by photo- and thermal-processes, allowing the resulting vacancies to provide a carrier doping effect in the TMD materials.[39,40]

### 3.3. Direct imaging of Schottky barrier position via SGM

To confirm the position of the potential barrier in device C, we performed SGM measurements as shown in Fig. 4. If a Schottky barrier exists at the contact region, the barrier height is modulated by the electrostatic interaction from the SGM tip when the biased tip situates just on the potential barrier. Consequently, the current across the barrier is affected (increased or decreased) and then the change of the current value appears as the SGM response. For negative gate voltage conditions ($V_{BG}$ = –10 V, p-type FET regime), a positive and a negative S-D bias were applied in Figs. 4(a) and 4(b), respectively. On the other hand, for positive gate voltage condition ($V_{BG}$ = 10 V, n-type FET regime), a positive and a negative S-D bias were applied in Figs. 4(d) and 4(e), respectively. The dark or the bright responses correspond to an increase or decrease of current across the channel depending on the carrier polarity and the tip bias voltage. Explaining the dark response (decrease of $I_{SD}$) in Fig. 4(a), for example, the bias condition during the SGM imaging corresponds to the transmission curve of the blue line in the green region in Fig. 3(d). Since the SGM response is in essence a from of gate-voltage response, it relates to the derivative of the gate-dependence slope (d$I_{SD}$/d$V_{BG}$) and therefore appears as negative. More precisely, when the positively biased tip situates on the position of the Schottky barrier at the interface of Pd/MoTe$_2$, as schematically shown in Fig. 4(c), the electric field from the tip lifts the local-barrier height. Consequently, the decrease of current value ($\Delta I_{SD}$) from the normal condition without the tip appears as the negative (dark) response in the SGM image. In all cases, the SGM response appears along the Pd/MoTe$_2$ interface suggesting the Schottky barrier formation arising from Fermi level pinning of MoTe$_2$ band as schematically shown in Fig. 4(c). On the other hand, no SGM response is observed at the interface between the MoTe$_2$ and the LIM in any condition suggesting achievement of universal Ohmic contact for both p- and n-type MoTe$_2$. Interestingly, even if the carriers are injected from the LIM contact, a thermionic-potential barrier (so-called diffusion potential) at the Pd/MoTe$_2$ interface also hinders the carrier emission from the channel in the low bias condition as schematically shown in Fig. 4(f).[23] In this case, electron injection from the laser-induced heavily-p-doped region to the n-type MoTe$_2$ channel could be achieved by a band to band tunneling (BTBT) regime. Note that the appearance of bright spots around the left electrode in Fig. 4(b) is not important since they arise from the leakage current between biased tip and the electrode during the luster scan of the biased tip.

## 4. Conclusions

In conclusion, we have confirmed the presence of a phase transition from semiconducting MoTe$_2$ to a laser-induced metal, due to photo-thermal decomposition that arises under high-density laser irradiation (of more than 0.3 MW/cm$^2$). The LIM phase includes pure Te and shows metallic behavior whose resistivity is 5 × 10$^{-6}$ Ωcm at room temperature. From transport measurements and SGM observations of the Pd/2H-MoTe$_2$/LIM FET device, the Pd/MoTe$_2$ interface was found to have a Schottky barrier, while there is no potential barrier at MoTe$_2$/LIM interface which therefore forms an Ohmic contact. The LIM contact is a universal method for achieving an Ohmic contact for both p- and n-type MoTe$_2$.

**Supporting information**

See the supplementary material for laser power density dependent phase transition, detailed setup for ambient scanning gate microscopy, EDS analysis of LIM phase, estimation of temperature during laser irradiation by using *h*-BN, density functional theory calculation of band structure of Te, and metallic behavior of LIM phase.

**Acknowledgements**

The authors appreciate fruitful discussions with Prof. Gil-ho Kim of SKKU, Korea. This work was supported by the JSPS KAKENHI Grant Numbers 16H00899 and 18H01812, and Chiba University VBL. We also thank JASCO Corp. for assistance with Raman mapping measurement in Fig. 2.

**FIGURES**

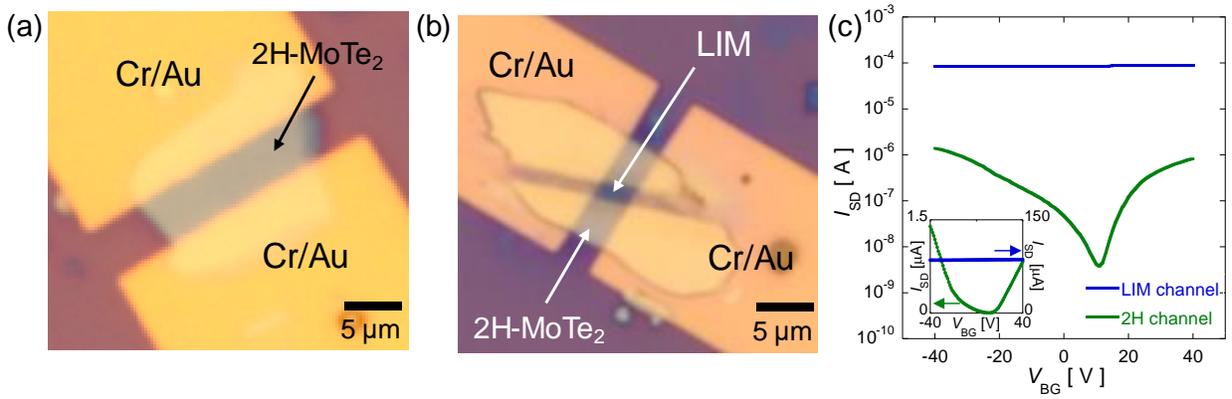

**Figure 1.** (a) Optical microscope image of device A; 2H-MoTe$_2$ FET fabricated on SiO$_2$/p$^{++}$Si substrate. (b) Optical microscope image of device B; 2H-MoTe$_2$ FET with LIM line. (c) $V_{BG}$ dependence of device A (green line) and device B (blue line). The inset shows the same data on a linear scale. $V_{SD}$ of 0.1 V was applied for both devices.

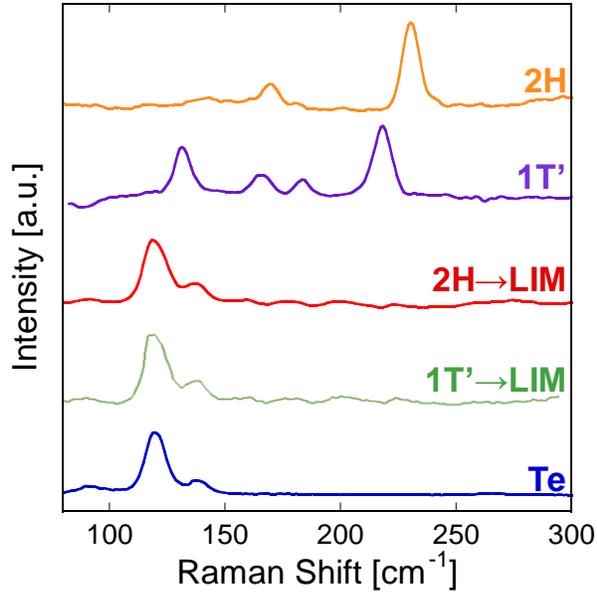

**Figure 2.** Raman spectra. From top to bottom: 2H-MoTe$_2$, thermally grown 1T′-MoTe$_2$, LIM phase from 2H, LIM phase from 1T′, and Te crystal.

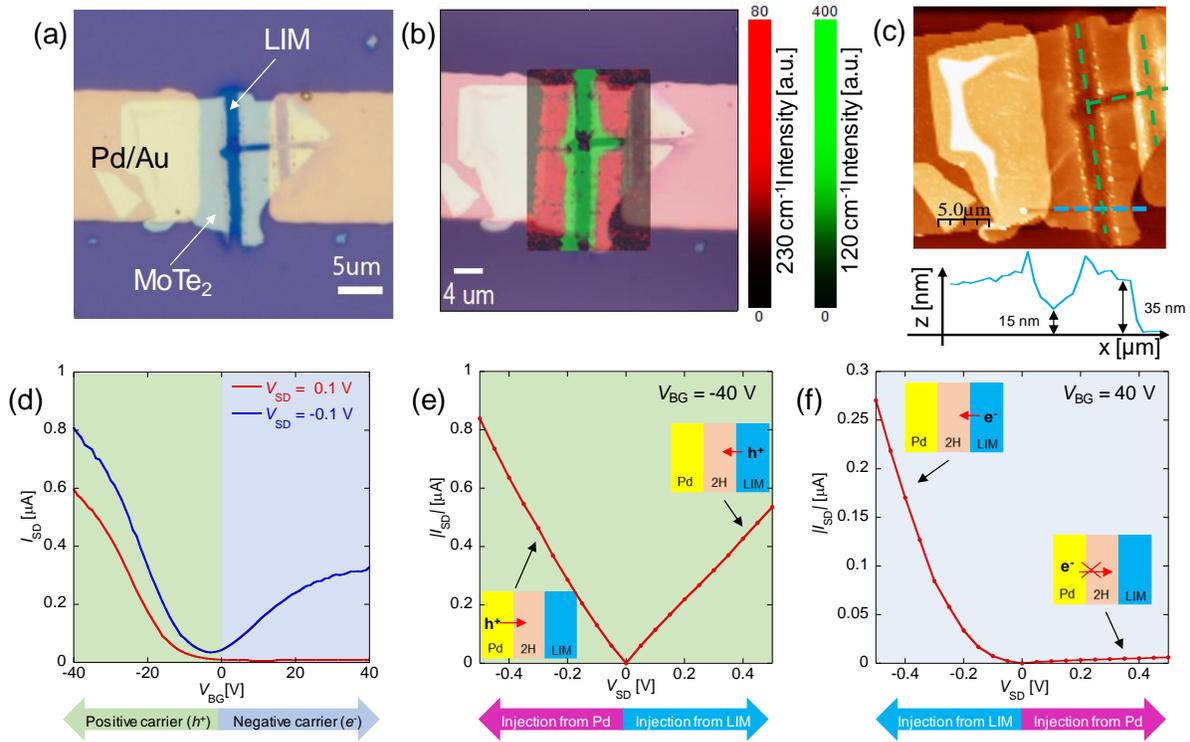

**Figure 3.** (a) Optical microscope image of device C; Pd/2H-MoTe$_2$/LIM junction FET device. (b) Raman spectroscopy map of the device. The green color indicates the main peak of the LIM phase and the red color indicates the 2H phase of MoTe$_2$. (c) AFM topographic image. The line-profile indicates the height from the SiO$_2$ plane along the blue colored dotted line. The green dashed lines indicate the laser irradiated regions. (d) $V_{BG}$ dependences of $I_{SD}$ at different bias condition. The blue and the red curves were obtained when the LIM contact is biased –0.1 V and 0.1 V, respectively. (e) $V_{SD}$ dependence of $I_{SD}$ at $V_{BG}$ = –40 V. (f) $V_{SD}$ dependence of $I_{SD}$ at $V_{BG}$ = 40 V. The insets show schematic diagrams of the carrier type and the direction of the injection into the channel in each bias condition.

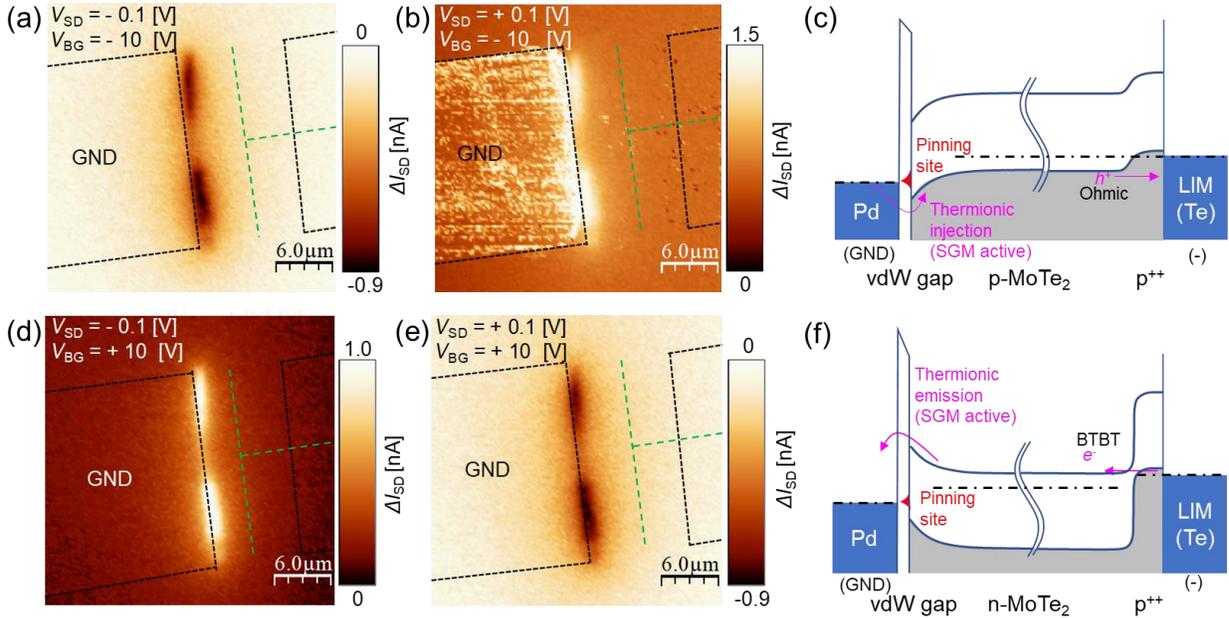

**Figure 4.** SGM response maps of device C at (a) $V_{SD} = -0.1$ V, $V_{BG} = -10$ V, (b) $V_{SD} = 0.1$ V, $V_{BG} = -10$ V. (c) Schematic band structure of p-type and negative bias condition corresponding to (a). (d) $V_{SD} = -0.1$ V, $V_{BG} = 10$ V, and (e) $V_{SD} = 0.1$ V, $V_{BG} = 10$ V. The color bars indicate the current variation in each image. Every image was taken under the condition of $V_T = 5$ V and interleave height of 100 nm. (f) Schematic band structure of n-type and negative bias condition corresponding to (d). Black dotted lines and green dotted ones in the SGM images indicate the position of the Pd/Au electrodes and LIM regions, respectively.